\documentclass[longauth]{aa}

\usepackage[varg]{txfonts}

\usepackage{graphicx}
\usepackage{epsfig}

\usepackage{hyperref}
\usepackage{natbib}
\bibpunct{(}{)}{;}{a}{}{,} 

\usepackage{amsmath}    
\usepackage{amssymb}    
\usepackage{bm}         

\usepackage{multicol}        
\usepackage{threeparttable}
\usepackage{tabularx}
\usepackage{float}
\usepackage{longtable}
\usepackage{supertabular}

\usepackage{color}

\usepackage{textcomp}

\newcommand{\pname}{HD\,219666\,b}
\newcommand{\sname}{HD\,219666}

\newcommand\tess{\emph{\it TESS}}
\newcommand\gaia{\emph{{\it Gaia}}}
\newcommand\tycho{\emph{{\it Tycho}}}
\newcommand\twomass{\emph{{\it 2MASS}}}

\newcommand{\ms}{$\mathrm{m\,s^{-1}}$}
\newcommand{\kms}{$\mathrm{km\,s^{-1}}$}

\newcommand\vsini{$v$\,sin\,$i_\star$}

\newcommand\teff{$T_{\rm eff}$}

\newcommand\logg{log\,{\it g$_\star$}}

\newcommand{\smass}[1][$M_{\odot}$]{$0.92 \pm 0.03$~#1} 
\newcommand{\sradius}[1][$R_{\odot}$]{$1.03 \pm 0.03$~#1}
\newcommand{\stemp}[1][$\mathrm{K}$]{$5527\pm65$~#1}
\newcommand{\Tzerob}[1][days]{$8329.1996 \pm 0.0012$~#1} 
\newcommand{\Pb}[1][days]{$6.03607 _{-0.00063}^{+0.00064}$~#1}

\newcommand{\bb}[1][ ]{$0.838_{-0.013} ^ {+0.012}$~#1} 
\newcommand{\arb}[1][ ]{$13.27\pm0.39$~#1} 
\newcommand{\rrb}[1][ ]{$0.04192\pm0.00083$~#1} 
\newcommand{\kb}[1][${\rm m\,s^{-1}}$]{$6.17\pm0.46$~#1} 
\newcommand{\mpb}[1][$M_{\oplus}$]{$16.6\pm1.3$~#1} 
\newcommand{\rpb}[1][$R_{\oplus}$]{$4.71\pm0.17$~#1}

\newcommand{\ib}[1][deg]{$86.38\pm0.15$#1} 
\newcommand{\ab}[1][AU]{$0.06356\pm0.00265$#1} 
\newcommand{\denpb}[1][${\rm g\,cm^{-3}}$]{$0.87_{-0.11}^{+0.12}$~#1} 
\newcommand{\Teqb}[1][K]{$1073\pm20$~#1} 
\newcommand{\ttotb}[1][hours]{$2.158\pm0.034$~#1} 
 
\newcommand{\qone}[1][]{$0.33\pm0.10 $#1} 
\newcommand{\qtwo}[1][]{$0.20\pm0.10 $#1}

\newcommand{\HARPS}[1][${\rm m\,s^{-1}}$]{$-20.0976\pm0.0004$~#1} 
\newcommand{\jHARPS}[1][${\rm m\,s^{-1}}$]{$1.04 _{-0.47}^{+0.48}$~#1} 

\newcommand\water{H$\sb{2}$O}
\newcommand\methane{CH$\sb{4}$}
\newcommand\carbdiox{CO$\sb{2}$}
\newcommand\ammonia{NH$\sb{3}$}
\newcommand\molhyd{H$\sb{2}$}

\begin{document}

\title{HD\,219666\,b: A hot-Neptune from \textit{TESS} Sector 1 
\thanks{Based on observations made with the 3.6m-ESO telescope at La Silla observatory under ESO programmes IDs 1102.C-0923 (PI: Gandolfi) and 1102.C-0249 (PI: Armstrong).}}

\author{M.~Esposito\inst{1}
    \and D.\,J.~Armstrong\inst{2,3}
    \and D.~Gandolfi\inst{4}
    \and V.~Adibekyan\inst{5}
    \and M.~Fridlund\inst{6,7}
    \and N.\,C.~Santos\inst{5,8}
    \and J.\,H.~Livingston\inst{9}
    \and E.~Delgado Mena\inst{5}
    \and L.~Fossati\inst{10}
    \and J.~Lillo-Box\inst{11}
    \and O.~Barrag\'an\inst{4}
    \and D.~Barrado\inst{12}
    \and P.\,E.~Cubillos\inst{10}
    \and B.\,Cooke\inst{2,3}
    \and A.\,B.~Justesen\inst{17}
    \and F.~Meru\inst{2,3}
    \and R.\,F.~D\'iaz\inst{13,14}
    \and F.~Dai\inst{22,15}
    \and L.\,D.~Nielsen\inst{16}
    \and C.\,M.~Persson\inst{6}
    \and P.\,J.~Wheatley\inst{2,3}
    \and A.\,P.~Hatzes\inst{1}
    \and V.~Van~Eylen\inst{15} 
    \and M.\,M.~Musso\inst{4}
    \and R.~Alonso\inst{18,19}
    \and P.\,G.~Beck\inst{18,19}
    \and S.\,C.\,C.~Barros\inst{5}
    \and D.~Bayliss\inst{2,3}
    \and A.\,S.~Bonomo\inst{35}
    \and F.~Bouchy\inst{16}
    \and D.\,J.\,A.~Brown\inst{2,3}
    \and E.~Bryant\inst{2,3}
    \and J.~Cabrera\inst{20}
    \and W\,.D.~Cochran\inst{21}
    \and S.~Csizmadia\inst{20}
    \and H.~Deeg\inst{18,19}
    \and O.~Demangeon\inst{5}
    \and M.~Deleuil\inst{23}
    \and X.~Dumusque\inst{16}
    \and P.~Eigm\"uller\inst{20}
    \and M.~Endl\inst{21}
    \and A.~Erikson\inst{20}
    \and F.~Faedi\inst{2,3}
    \and P.~Figueira\inst{11,5}
    \and A.~Fukui\inst{24}
    \and S.~Grziwa\inst{25}
    \and E.\,W.~Guenther\inst{1}
    \and D.~Hidalgo\inst{18,19}
    \and M.~Hjorth\inst{17}
    \and T.~Hirano\inst{26}
    \and S.~Hojjatpanah\inst{5,8}  
    \and E.~Knudstrup\inst{17}
    \and J.~Korth\inst{25}
    \and K.\,W.\,F.~Lam\inst{28}
    \and J.~de~Leon\inst{9}
    \and M.\,N.~Lund\inst{17}
    \and R.~Luque\inst{18,19}
    \and S.~Mathur\inst{18,19}
    \and P.~Monta\~n\'es Rodr\'iguez\inst{18,19}
    \and N.~Narita\inst{9,24,29,18,36}
    \and D.~Nespral\inst{18,19}
    \and P.~Niraula\inst{27}
    \and G.~Nowak\inst{18,19}
    \and H.\,P.~Osborn\inst{23}
    \and E.~Pall\'e\inst{18,19}
    \and M.~P\"atzold\inst{25}
    \and D.~Pollacco\inst{2,3}
    \and J.~Prieto-Arranz\inst{18,19}
    \and H.~Rauer\inst{20,28,31}
    \and S.~Redfield\inst{30}
    \and I.~Ribas\inst{32,33}
    \and S.\,G.~Sousa\inst{5}
    \and A.\,M.\,S.~Smith\inst{20}
    \and M.~Tala-Pinto\inst{34}
    \and S.~Udry\inst{16}
    \and J.\,N.~Winn\inst{15}
   }
    
\institute{Th\"uringer Landessternwarte Tautenburg, Sternwarte 5, D-07778 Tautenburg, Germany \email{mesposito@tls-tautenburg.de}
     \and Department of Physics, University of Warwick, Gibbet Hill Road, Coventry, CV4 7AL
     \and Centre for Exoplanets and Habitability, University of Warwick, Gibbet Hill Road, Coventry, CV4 7AL
     \and   Dipartimento di Fisica, Universit\`a degli Studi di Torino, via Pietro Giuria 1, I-10125, Torino, Italy 
     \and Instituto de Astrofísica e Ciências do Espaço, Universidade do Porto, CAUP, Rua das Estrelas, 4150-762 Porto, Portugal 
     \and Department  of  Space,  Earth  and  Environment,  Chalmers  University  of  Technology,  Onsala  Space  Observatory,  439  92  Onsala, Sweden
     \and Leiden Observatory, University of Leiden, PO Box 9513, 2300 RA Leiden, The Netherlands
     \and Departamento de Física e Astronomia, Faculdade de Ciencias, Universidade do Porto, Rua Campo Alegre, 4169-007 Porto, Portugal 
     \and Department of Astronomy, Graduate School of Science, The University of Tokyo, Hongo 7-3-1, Bunkyo-ku, Tokyo, 113-0033, Japan
     \and Space Research Institute, Austrian Academy of Sciences, Schmiedlstrasse 6, A-8041 Graz, Austria
     \and European Southern Observatory, Alonso de Cordova 3107, Vitacura, Santiago, Chile
     \and Depto. de Astrofísica,  Centro de Astrobiología (INTA-CSIC), Campus ESAC (ESA)Camino Bajo del Castillo s/n 28692 Villanueva de la Cañada, Spain 
     \and Universidad de Buenos Aires, Facultad de Ciencias Exactas y Naturales. Buenos Aires, Argentina
     \and CONICET - Universidad de Buenos Aires. Instituto de Astronom\'ia y F\'isica del Espacio (IAFE). Buenos Aires, Argentina
     \and Department of Astrophysical Sciences, Princeton University, 4 Ivy Lane, Princeton, NJ, 08544, USA
     \and Geneva Observatory, University of Geneva, Chemin des Mailettes 51, 1290 Versoix, Switzerland
     \and Stellar Astrophysics Centre, Deparment of Physics and Astronomy, Aarhus University, Ny Munkegrade 120, DK-8000 Aarhus C, Denmark
     \and Instituto de Astrofísica de Canarias, C/ Vía Láctea s/n, E-38205, La Laguna, Tenerife, Spain
     \and Departamento de Astrofísica, Universidad de La Laguna, E-38206, Tenerife, Spain 
     \and Institute of Planetary Research, German Aerospace Center, Rutherfordstrasse 2, 12489 Berlin, Germany
     \and Department of Astronomy and McDonald Observatory, University of Texas at Austin, 2515 Speedway, Stop C1400, Austin, TX 78712, USA
     \and Department of Physics and Kavli Institute for Astrophysics and Space Research, MIT, Cambridge MA 02139 USA
     \newpage
     \and Aix Marseille Univ, CNRS, CNES, LAM, Marseille, France
     \and National Astronomical Observatory of Japan, NINS, 2-21-1 Osawa, Mitaka, Tokyo 181-8588 Japan
     \and Rheinisches Institut für Umweltforschung an der Universität zu Köln, Aachener Strasse 209, D-50931 Köln
     Germany
     \and Department of Earth and Planetary Sciences, Tokyo Institute of Technology, Meguro-ku, Tokyo Japan 
     \and Department of Earth, Atmospheric and Planetary Sciences, MIT, 77 Massachusetts Avenue, Cambridge, MA 02139
     \and Zentrum für Astronomie und Astrophysik, Technische Universität Berlin, Hardenbergstr. 36 D-10623 Berlin
     Germany
     \and Astrobiology Center, NINS, 2-21-1 Osawa, Mitaka, Tokyo, 181-8588, Japan
     \and Astronomy Department and Van Vleck Observatory, Wesleyan University, Middletown, CT 06459, USA
     \and Institut für Geologische Wissenschaften, Freie Universität Berlin, Malteserstr. 74–100 D-12249 Berlin
     Germany
     \and Institut de Ciències de l’Espai (ICE, CSIC), Campus UAB, Bellaterra Spain
     \and Institut d’Estudis Espacials de Catalunya (IEEC), Barcelona, Spain
     \and Landessternwarte K\"{o}nigstuhl, Zentrum f\"{u}r Astronomie der Universit\"{a}t Heidelberg, K\"{o}nigstuhl 12, 69117 Heidelberg, Germany 
     \and INAF – Osservatorio Astrofisico di Torino, Via Osservatorio
     20, I-10025 Pino Torinese, Italy
     \and JST, PRESTO, 7-3-1 Hongo, Bunkyo-ku, Tokyo, 113-0033, Japan
    %
    }

\date{Received <date> /
      Accepted <date>}
      
\abstract {We report on the confirmation and mass determination of a transiting planet orbiting the old and 
inactive G7 dwarf star HD\,219666 ($M_{\star}$=\smass, $R_{\star}$=\sradius, $\tau_{\star}$=10\,$\pm$\,2 Gyr). 
With a mass of $M_{\rm b}$\,=\,\mpb, a radius of $R_{\rm b}$\,=\,\rpb, and an orbital period of $P_\mathrm{orb}$\,$\simeq$\,6\,days, HD\,219666\,b is a new member of a 
rare class of exoplanets: the hot-Neptunes. The Transiting Exoplanet Survey Satellite (\tess) observed HD\,219666 (also known as TOI-118) in its Sector 1 
and the light curve shows four transit-like events, equally spaced in time. We confirmed the planetary nature of the candidate by gathering precise radial-velocity 
measurements with the High Accuracy Radial velocity Planet Searcher (HARPS) at ESO\,3.6m. We used the co-added HARPS spectrum to derive the host star fundamental parameters 
(\teff\,=\,\stemp, \logg\,=\,4.40\,$\pm$\,0.11 (cgs), [Fe/H]\,=\,0.04\,$\pm$\,0.04\,dex, $\log R^{\prime}_{\rm HK}$\,=\,$-$5.07\,$\pm$\,0.03), as well as the abundances 
of many volatile and refractory elements. The host star brightness (V=9.9) makes it suitable for further characterisation by means of in-transit spectroscopy. 
The determination of the planet orbital obliquity, along with the atmospheric metal-to-hydrogen content and thermal structure could provide us with important clues on 
the formation mechanisms of this class of objects.} 
  
           
\titlerunning{The transiting hot-Neptune HD\,219666\,b}
\authorrunning{M. Esposito et al.}           
  
\maketitle





\section{Introduction}\label{Sec:Intro}

Following the success of the \textit{Kepler} space mission \citep{Borucki2016},  in April 2018 NASA launched a new satellite, the Transiting 
Exoplanet Survey Satellite \citep[\tess,][]{Ricker2015}. By performing a full-sky survey, \tess\ is expected to detect approximately 10\,000 transiting 
exoplanets (TEPs) \citep{Barclay2018,Huang2018a}. Most interestingly,  nearly 1000 of them will orbit host stars with magnitudes V$\lesssim$10\, (as of 
November 2018 there are  56 known TEPs around stars with V<10, only 13 of which have masses <\,20 M$_{\oplus}$, according to the NASA 
exoplanet archive\footnote{https://exoplanetarchive.ipac.caltech.edu/.}). Bright host stars are suitable for precise radial-velocity (RV) measurements 
that can lead to planet mass determinations down to a few Earth masses, and to estimates of the planet bulk density for TEPs. In-transit precise RVs  
also allow us to measure the planet orbital obliquity through the observation of the Rossiter-McLaughlin effect \citep[see, e.g.][]{Triaud2017}. High-S/N spectra 
are very much needed for transmission spectroscopy studies aimed at the detection of atomic and molecular species, and the characterisation of the thermal 
structure of planet atmospheres \citep{Snellen2010,Bean2013}.

\tess\ has a field of view of 24°\,$\times$\,96°, and will cover almost the full sky in 26 Sectors, each monitored for about 27 days. Full frame 
images (FFIs) are registered every 30 minutes, while for a selected sample of bright targets ($\sim$16\,000 per Sector) pixel sub-arrays are saved with a 
two-minute cadence. 
 The first \tess\, data  set  of  FFIs  from  Sectors  1  and  2  was  released  on December 6, 2018, and the \tess\, Science Office, supported by
the Payload Operations Centre at MIT, had already issued \tess\, data alerts for a number of transiting planet-host star candidates,
the so-called \tess\, objects of interest (TOIs). Preliminary two-minute cadence light curves and  target pixel files \citep{Twicken2018} are made publicly available 
for download at the MAST web site\footnote{Mikulski Archive for Space Telescopes, https://archive.stsci.edu/prepds/tess-data-alerts/.}.

Several \tess\, confirmed planets have already been announced: \object{$\pi$ Mensae c} (TOI-144), a super-Earth 
orbiting a V=5.65 mag  G0 V star \citep{Huang2018b,Gandolfi2018}; \object{HD\,1397\,b} (TOI-120), a warm giant planet around a V=7.8 mag sub-giant 
star \citep{Brahm2018,Nielsen2018}; \object{HD\,2685\,b} (TOI-135), a hot-Jupiter hosted by an early F-type star \citep{Jones2018}; and an ultra-short-period Earth-like planet 
around the M-dwarf star \object{LHS\,3844} \citep[TOI-136;][]{Vanderspek2018}. Here we report on the detection and mass determination of a Neptune-like 
planet ($M_{\mathrm b}\,\simeq\,16.6\,M_{\oplus}$,~$R_{\mathrm b}\,\simeq\,4.7\,R_{\oplus}$) on a $P_\mathrm{orb} \simeq 6$ day orbit around the 
bright (V=9.9) G7\,V star \object{HD\,219666} (TOI-118; Table~\ref{table:target}). 

The work presented here is part of the ongoing RV follow-up effort carried out by two teams, namely the \texttt{KESPRINT} 
consortium \citep[see, e.g.][]{Johnson2016, VanEylen2016, Dai2017, Gandolfi2017, Barragan2018, PrietoArranz2018} and the \texttt{NCORES} 
consortium \citep[see, e.g.][]{Armstrong2015, Lillo-Box2016, Barros2017, Lam2018, Santerne2018}. Both teams were recently awarded two large 
programs with the High Accuracy Radial velocity Planet Searcher (HARPS) spectrograph at the ESO-3.6m telescope to follow up \tess\ transiting planet candidates. The two consortia have joined forces 
to make better use of the instrument, optimise the scientific return of the available observing time, and tackle more ambitious planet detections and characterisations. 

This paper is organised as follows. Section \ref{Sec:Tess_Photometry} describes the \tess\, photometric data, our custom light-curve 
extraction and assessment of the light contamination factor. Section \ref{Sec:spectroscopy} reports on our spectroscopic follow-up observations, which were 
used to confirm the planetary nature of the transiting companion, and to derive the fundamental parameters and metal abundances of the 
host star (Section \ref{Sec:StellarParameters}). The joint analysis of transit light curves and RV data is 
described in Section \ref{Sec:JointAnalysis}. Finally we discuss our results in Section \ref{Sec:Discussion}.

\begin{table}
\caption{Main identifiers, coordinates, parallax, and optical and infrared
magnitudes of \sname.}
\label{table:target}
\centering
\begin{tabular}{lrr}
\hline\hline
\noalign{\smallskip}
Parameter & Value & Source \\
\noalign{\smallskip}
\hline
\noalign{\smallskip}
HD & 219666 &  \\
TIC ID & 266980320 & TIC  \\
TOI ID & 118       & \tess\ Alerts \\
\textit{Gaia} DR2 ID & 6492940453524576128 & \textit{Gaia} DR2 \tablefootmark{a} \\
RA (J2000)  & 23$^h$ 18$^m$ 13.630$^s$ &  \gaia\ DR2 \tablefootmark{a} \\
DEC (J2000) & -56° 54' 14.036'' & \gaia\ DR2 \tablefootmark{a} \\
$\mu_{\rm RA}$ [mas yr$^{-1}$] & 313.918 $\pm$ 0.039 & \gaia\ DR2 \tablefootmark{a}  \\
$\mu_{\rm DEC}$ [mas yr$^{-1}$] & -20.177 $\pm$ 0.043 & \gaia\ DR2  \tablefootmark{a} \\
$\pi$ [mas] & $10.590 \pm 0.028$ & \gaia\ DR2 \tablefootmark{a} \\
$B_\mathrm{T}$ & $10.785  \pm 0.027$ & \tycho-2 \tablefootmark{b} \\
$V_\mathrm{T}$ & $9.897   \pm 0.018$ & \tycho-2 \tablefootmark{b} \\
$G$            & $9.6496  \pm 0.0002 $ & \gaia\ DR2 \tablefootmark{a} \\
$G_\mathrm{BP}$& $10.0349 \pm 0.0009 $ & \gaia\ DR2 \tablefootmark{a} \\
$G_\mathrm{RP}$& $ 9.1331 \pm 0.0008 $ & \gaia\ DR2 \tablefootmark{a} \\
$J$            & $ 8.557  \pm 0.020  $ & 2MASS \tablefootmark{c} \\
$H$            & $ 8.254  \pm 0.042  $ & 2MASS \tablefootmark{c} \\
$K_s$          & $ 8.158  \pm 0.033  $ & 2MASS \tablefootmark{c} \\
$W1$(3.35 $\mu$m)   &  $8.080 \pm 0.023$  & WISE \tablefootmark{d}   \\
$W2$(4.6 $\mu$m)    &  $8.138 \pm 0.020$  & WISE \tablefootmark{d}   \\
$W3$(11.6 $\mu$m)   &  $8.100 \pm 0.021$  & WISE \tablefootmark{d}   \\
$W4$(22.1 $\mu$m)   &  $8.250 \pm 0.288$  & WISE \tablefootmark{d}   \\
\noalign{\smallskip}
\hline
\end{tabular}
\tablefoot{\tablefoottext{a}{\citet{GaiaDR2}.} \tablefoottext{b}{\citet{Hog2000}.} 
\tablefoottext{c}{\citet{Cutri2003}.} \tablefoottext{d}{\citet{Cutri2013}.} 
}
\end{table}

\section{\tess\ photometry}\label{Sec:Tess_Photometry}

\begin{figure}
\resizebox{\hsize}{!}{\includegraphics[trim={2mm 7.5mm 2mm 13mm},clip]{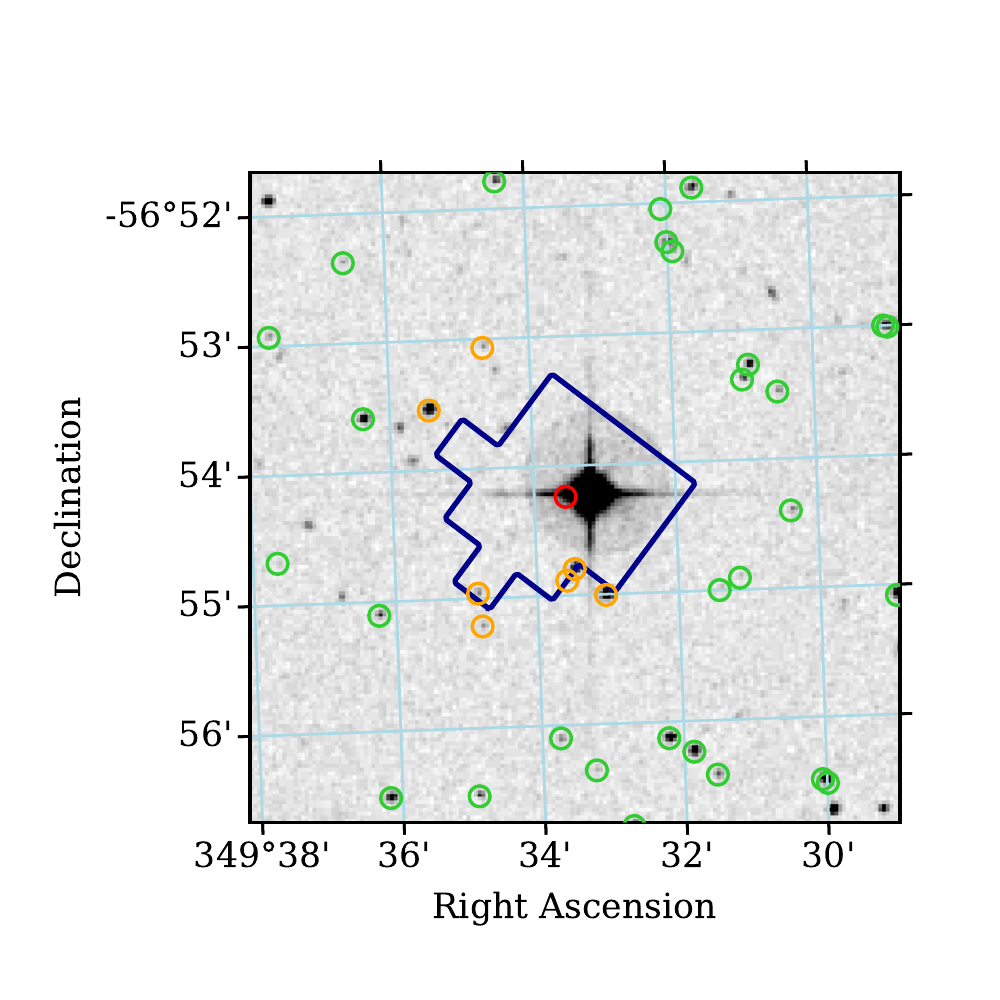}}
\caption{5\arcmin$\times$5\arcmin\ archival image taken in 1980 from the SERCJ survey, with the SPOC photometric aperture 
overplotted in blue (\tess\, pixel size is 21\arcsec), and the positions of \gaia\ DR2 sources (J2015.5) within 2\arcmin\ of \sname\ indicated 
by circles. \sname\ is in red, nearby sources contributing more than 1\% of their flux to the aperture are in orange (see Sect.~\ref{Subec:contam}), and other sources are in green.} 
\label{fig:tessaperture}
\end{figure}

HD\,219666 was observed by \tess\, in Sector 1 (CCD \#2 of Camera \#2) and falls in a region of the sky that will not be further 
visited by \tess. Sector 1 was monitored continuously  for $\sim$27.9~days, from 2018-07-25 (BJD$_\mathrm{TDB}$ = 2458325.29953) to 
2018-08-22 (BJD$_\mathrm{TDB}$ = 2458353.17886), with only a 1.14 day gap (from BJD$_\mathrm{TDB}$ = 2458338.52472 to BJD$_\mathrm{TDB}$ = 2458339.66500) when 
the satellite was repointed for data downlink.  In addition, between BJD$_\mathrm{TDB}$ = 2458347 and BJD$_\mathrm{TDB}$ = 2458350, the \tess\ light curve shows 
a higher noise level caused by the spacecraft pointing instabilities. The corresponding data-points were masked out and not included in the analysis presented in this paper. 

\subsection{Custom light-curve preparation}\label{Subec:lightcurve}

To check that the SPOC aperture is indeed an optimal choice, we extracted a series of light curves from the pixel data using contiguous sets of pixels 
centred on \sname. We first computed the 50$^\mathrm{th}$ to 95$^\mathrm{th}$ percentiles (in 1\% steps) of the median image, and then selected pixels 
with median counts above each percentile value to form each aperture. We then computed the 6.5 hour combined differential photometric 
precision (CDPP) \citep{Christiansen2012} of the light curve resulting from each of these apertures, and we found that the aperture that minimised the 
CDPP was slightly larger than the SPOC aperture shown in Fig.~\ref{fig:tessaperture}. However, we opted to use the PDCSAP light curve produced from 
the SPOC aperture, which has lower levels of systematic noise as a result of the processing performed by the SPOC pipeline \citep{Ricker2018}. 

\begin{figure*}
\includegraphics[width=18cm]{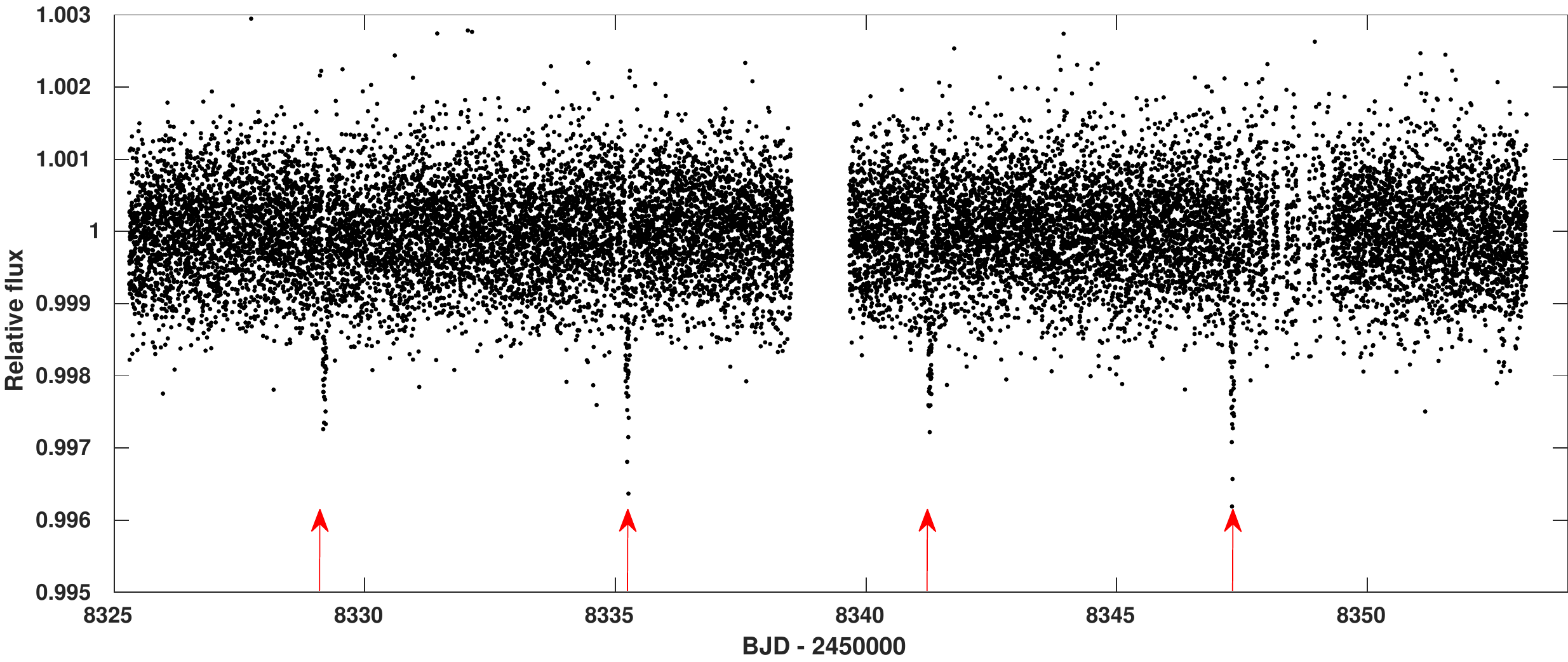}
\caption{The \tess\ light curve of HD\,219666. The red arrows point to the four planet-transit occurrences.}
\label{fig:TESS_lc}
\end{figure*}

The median-normalised light curve that we used in our analysis is shown in Fig.\,\ref{fig:TESS_lc}. 

\subsection{Limits on photometric contamination}\label{Subec:contam}

To investigate the possibility of contaminating flux from nearby stars within the SPOC photometric aperture, we compared 
the \gaia\ DR2 \citep{GaiaDR2} sources with the aperture and an archival image of \sname\ from 
the SERC-J survey\footnote{Available at \url{http://archive.stsci.edu/cgi-bin/dss_form}.}. To do so, we executed a query centred on 
the coordinates of \sname\ from the \tess\ Input Catalog\footnote{Available at \url{https://mast.stsci.edu/portal/Mashup/Clients/Mast/Portal.html}.} \citep[TIC;][]{Stassun2018} 
using a search radius of 3\arcmin. The archival image, taken in 1980, shows \sname\ to be offset from its current position 
by $\sim$4.8\arcsec. The proper motion is not sufficient to completely rule out chance alignment with a background source, but such an alignment 
with a bright source is qualitatively unlikely. We also note the non-detection by \gaia\ of any other sources 
within $\sim$30\arcsec\ of \sname. Fig.~\ref{fig:tessaperture} shows \gaia\, DR2 source positions overplotted on the archival image, along with 
the SPOC photometric aperture. Using a 2D Gaussian profile with a FWHM of $\sim$25\arcsec\ to approximate the \tess\, point spread 
function (PSF), and a negligible difference between the $G_\mathrm{RP}$ and $T$ bandpasses, we found that the transit depth of \sname\ should be 
diluted by no more than 0.1\%, even considering partial flux contributions from nearby stars outside the aperture. Furthermore, we found 
that \sname\ is the only star in or near the aperture that is bright enough to be the source of the transit signal, given the observed depth 
and assuming a maximum eclipse depth~of~100\%.

\section{HARPS observations}\label{Sec:spectroscopy}

\begin{figure}
\resizebox{\hsize}{!}{\includegraphics{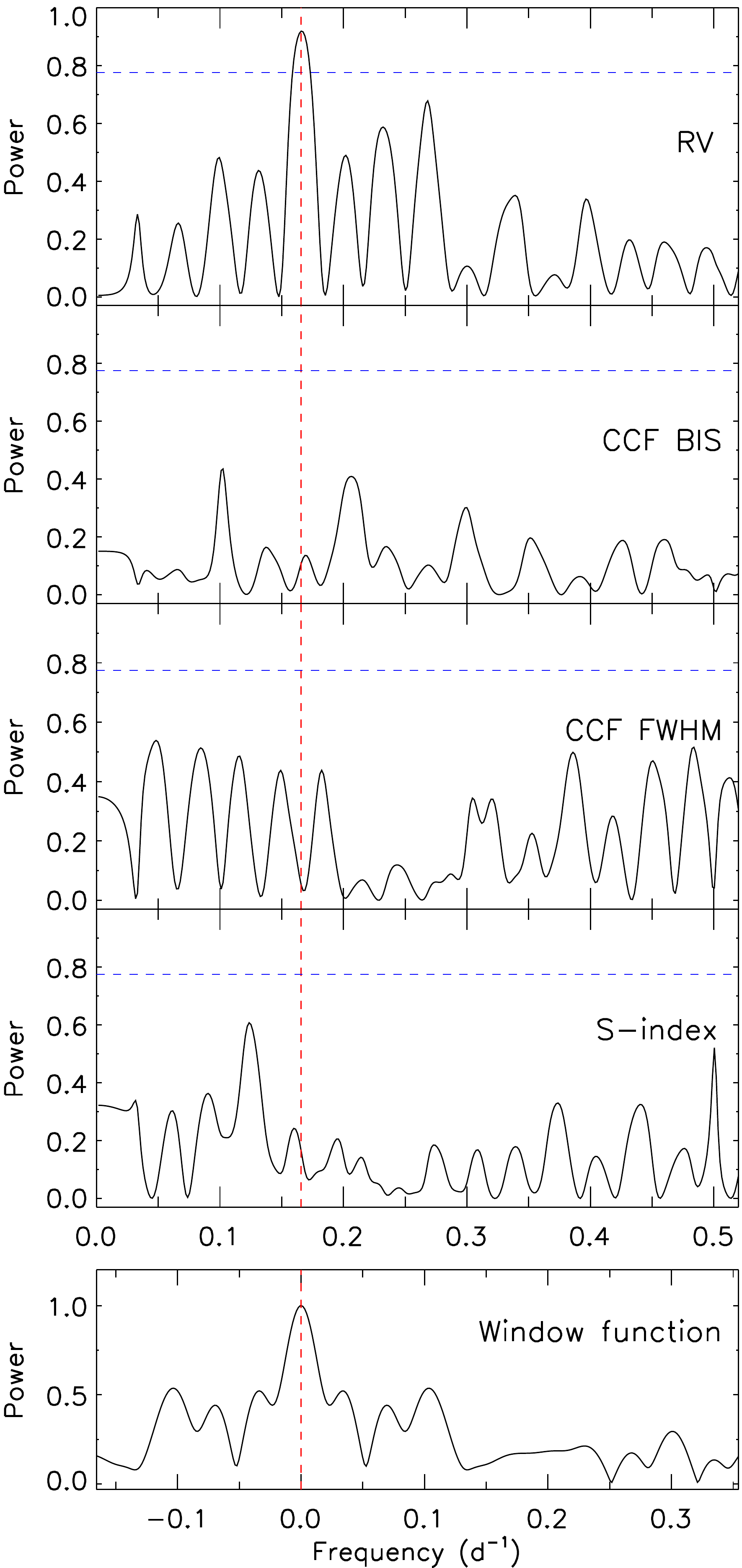}}
\caption{ Generalised Lomb-Scargle periodogram of the HARPS RVs (first panel), the CCF BIS and FWHM (second  and third panel, respectively), 
the Ca {\sc ii} H and K S-index (fourth panel), and of the window function (fifth panel). 
The vertical dashed red line marks the frequency of the transit 
signal. The horizontal dashed blue line marks the FAP\,=\,0.1\% level.}
\label{fig:GLS_DopplerData}
\end{figure}

We acquired 21 high-resolution ($R$\,$\approx$\,$115\,000$) spectra of \sname\ with the   
HARPS spectrograph \citep{Mayor2003} mounted at the ESO-3.6m telescope of La Silla observatory (Chile). The observations were performed 
between 02 October and 05 November 2018 UTC, as part of the large observing programmes 1102.C-0923 (PI: Gandolfi) and 1102.C-0249 (PI: Armstrong). We reduced 
the data using the dedicated HARPS Data Reduction Software (DRS) and extracted the RVs by cross-correlating the echelle 
spectra with a G2 numerical mask \citep{Baranne1996, Pepe2002, Lovis2007}. Table~\ref{Table:HARPS} lists the HARPS RVs and their uncertainties, along with 
the BIS and FWHM of the cross-correlation function (CCF), the Ca\,{\sc ii}\,H and K Mount-Wilson S-index, and S/N per pixel at 5500\,\AA.

The generalised Lomb-Scargle \citep[GLS;][]{Zechmeister2009} periodogram of the HARPS RV measurements (Fig.~\ref{fig:GLS_DopplerData}, first panel) shows 
a significant peak at the frequency of the transit signal ( $f_1$\,=\,0.166 d$^{-1}$; vertical dashed red line), with a false alarm 
probability\footnote{Computed following the Monte Carlo bootstrap method described in \citet{Kuerster1997}.} (FAP) lower than 0.1\,\% (horizontal dashed blue line). 
The peak has no counterpart in the periodograms of the activity indicators, as shown in the second, third, and fourth panels of Fig.~\ref{fig:GLS_DopplerData}. This provides 
strong evidence that the signal detected in our Doppler data is induced by an orbiting companion and confirms the presence of the transiting 
planet with a period of about 6 days. The periodogram of the RV measurements shows additional peaks symmetrically distributed to the left and 
right of the dominant frequency. We interpreted these peaks as aliases of the orbital frequency, as shown by the position of the peaks in 
the periodogram of the window function (Fig.~\ref{fig:GLS_DopplerData}, fifth panel).

\section{Stellar fundamental parameters}\label{Sec:StellarParameters}

The determination of the stellar parameters from the spectrum of the host star is crucial in order to derive the planetary parameters from transit 
and RV data. The three most important planetary parameters are the mass, $M_\mathrm{b}$, the radius $R_\mathrm{b}$, and the age $\tau_\mathrm{b}$, all of 
them only derivable with knowledge of the same parameters for the host star, $M_{\star}$, $R_{\star}$, and $\tau_{\star}$. Therefore we have used two independent 
methods in order to determine the stellar parameters with the highest degree of confidence available today. To this aim, we used the co-added HARPS spectrum, which 
has a S/N per pixel of $\sim$300 at 5500\,\AA. 

In one of the methods, we used version 5.22 of the Spectroscopy Made Easy (SME) code   \citep{Valenti1996, Valenti2005, Piskunov2017}. The SME code calculates 
synthetic spectra, using a grid of stellar models and a set of initial (assumed) fundamental stellar parameters and fits the result to the observed 
high-resolution spectrum with a chi-square minimisation procedure. The code contains a large library of different 1D and 3D model grids. In our analysis of the 
co-added HD\,219666 HARPS spectrum, we used the ATLAS12 model atmosphere grid \citep{Kurucz2013}. This is a set of 1D models applicable to solar-like stars. The observed 
spectral features that we fit are sensitive to the different photospheric parameters, including the effective temperature \teff, metallicity [M/H], surface 
gravity \logg, micro- and macro-turbulent velocities $v_{\mathrm mic}$ and $v_{\mathrm mac}$, and the projected rotational velocity \vsini. In order to minimise 
the number of free parameters we adopted the calibration equation of \citet{Bruntt2010} to estimate $v_{\mathrm mic}$ and we fitted many isolated and unblended 
metal lines to determine \vsini.
 
We used several different observed spectral features as indicators of each fundamental stellar parameter. The \teff\ was primarily determined by 
fitting the wings of Balmer lines, which for solar-type stars are almost totally dependent on the temperature and weakly dependent on gravity 
and metallicity \citep{Fuhrmann1993}. The surface gravity \logg\ was determined by fitting the line profiles of the Ca\,{\sc i} lines at 6102, 6122, 6162, and 6439 \AA, and the 
profiles of the Mg\,{\sc i} triplet at 5160-5185 \AA. Results were then checked by fitting also the line wings of the sodium doublet 
at 5896 and 5890 \AA\ using a sodium abundance determined from a number of fainter lines. In this case all three ions provided the same value for \logg. Using this 
method we derived an effective temperature \teff\,=\,5450\,$\pm$\,70\,K, surface gravity \logg\,=\,4.35\,$\pm$\,0.06 (cgs), iron 
content of [Fe/H]\,=\,+0.06\,$\pm$\,0.03 dex, calcium content of [Ca/H]\,=\,0.12\,$\pm$\,0.05 dex, magnesium [Mg/H]\,=\,0.18\,$\pm$\,0.10 dex, and 
sodium [Na/H]\,=\,0.15\,$\pm$\,0.01 dex. The $v_{\mathrm mic}$ used was 0.9\,$\pm$\,0.1~\kms, and we found \vsini\,=\,2.2\,$\pm$\,0.8~\kms\ and $v_{\rm mac}$\,=\,2.8\,$\pm$\,0.9~\kms. 

In an independent analysis, stellar atmospheric parameters (\teff, \logg, $v_{\mathrm mic}$, and [Fe/H]) and respective error bars were derived using the 
methodology described in \citet{Sousa-14} and \citet{Santos-13}. Briefly, we made use of the equivalent widths (EWs) of 224 Fe\,{\sc i} and 35 Fe\,{\sc ii} lines, as 
measured in the combined HARPS spectrum of HD\,219666 using the ARES v2 code\footnote{The last version of the ARES code (ARES v2) can be downloaded 
at \url{http://www.astro.up.pt/$\sim$sousasag/ares}.} \citep{Sousa-15}, and we assumed ionisation and excitation equilibrium. The process makes use of 
a grid of ATLAS model atmospheres \citep{Kurucz-93} and the radiative-transfer code MOOG \citep{Sneden-73}. This method 
provides effective temperatures in excellent agreement with values derived using the infrared flux method that are independent of the derived 
surface gravity. The resulting values are \teff\,=\,5527\,$\pm$\,25 K, \logg = 4.34$\pm$0.04 (cgs), $v_{\rm mic}$\,=\,0.90\,$\pm$\,0.04~\kms, and [Fe/H] = 0.04\,$\pm$\,0.02 dex. The 
surface gravity corrected for the systematic effects discussed in \citet{Mortier-13} has a value of \logg = 4.40$\pm$0.04 (cgs). 

The two sets of spectroscopic parameters obtained using the two independent methods described above are in good agreement. While we have no reason to prefer 
one method over the other, in the following analyses we adopted the values derived using the EW method. We stress that the quoted uncertainties 
are internal error bars that do not account for the choice of spectral lines and/or atmospheric models. Following \citet{Sousa2011}, we accounted for systematic 
effects by quadratically adding 60\,K, 0.1 (cgs), and 0.04 dex to the nominal uncertainty of the effective temperature, surface gravity, and 
iron content, respectively. The adopted values of \teff\,=\,5527\,$\pm$\,65 K, \logg = 4.40\,$\pm$\,0.11 (cgs), and [Fe/H] = 0.04\,$\pm$\,0.04 dex are listed in Table~\ref{table:starspecpar}.

Stellar abundances of the elements were also derived using the same tools and models as for stellar parameter determination, as well as using the classical 
curve-of-growth analysis method, assuming local thermodynamic equilibrium (LTE). Although the EWs of the spectral lines were automatically measured with ARES, for the elements 
with only two or three lines available we performed careful visual inspection of the EWs. For the derivation of chemical abundances of refractory elements  
we closely followed the methods described in \citet[e.g.][]{Adibekyan-12,Adibekyan-15,Delgado-17}. Abundances of the volatile elements O and C were derived 
following the method of \citet{Delgado-10} and \citet{Bertrandelis-15}. Since the two spectral lines of oxygen are usually weak and the 6300.3\AA{} line is blended with 
Ni and CN lines, the EWs of these lines were manually measured with the task \texttt{splot} in IRAF. We noticed that for several individual spectra of 
the star, the 6300\AA{} region was contaminated by the telluric [OI] emission line. We excluded these contaminated spectra when measuring the EW of the 
stellar oxygen line at 6300.3\AA{}. Lithium and sulfur abundances were derived by performing spectral synthesis with MOOG \citep{Delgado-14}. The final abundances 
of the elements are presented in Table~\ref{table:starspecpar}. It is worth noting that the abundances of Na, Mg, and Ca derived with this EW method are 
in agreement with the abundances obtained with the spectral fitting method. Perhaps it is also interesting to note that the star seems to be enhanced 
in several $\alpha$ elements (Mg, Si, Ti) and show under-abundance of some heavy elements (e.g. Ba and Y). Such a chemical composition is typical for 
the so-called high-$\alpha$ metal-rich stars first discovered by \citet[][]{Adibekyan-11, Adibekyan-13}. The origin of this population is not yet fully 
clear, but most probably these stars are migrators from the inner Galaxy \citep{Adibekyan-11, Anders-18}

We derived the stellar radius ($R_\star$) combining the \tycho\ $B_\mathrm{T}$, $V_\mathrm{T}$ magnitudes, the \gaia\ $G$, $G_\mathrm{BP}$, $G_\mathrm{RP}$ 
photometry, and \twomass\ $J$, $H$, $K_s$ magnitudes (see Table~\ref{table:target}) with our spectroscopic 
parameters (\teff, \logg, [Fe/H]; see Table~\ref{table:starspecpar}) and the \gaia' parallax \citep[10.590\,$\pm$\,0.028~mas,][ see Table~\ref{table:starspecpar}]{GaiaDR2}. We 
corrected the \gaia\ G photometry for the magnitude dependent offset using Eq.\,3 from \citet{Casagrande2018}, and adopted a minimum uncertainty 
of 0.01 mag for the \gaia\ magnitudes to account for additional systematic uncertainties in the \gaia\ photometry. We added 0.06~mas to the 
nominal \gaia's parallax to account for the systematic offset found by \citet{StassunTorres2018}, \citet{Riess2018}, and \citet{Zinn2018}. Following the method 
described in \citet{Gandolfi2008}, we found that the reddening along the line of sight to the star is consistent with zero and did not correct the 
apparent magnitudes. The bolometric correction for each band-pass was computed using the routine from \citet{Casagrande2018}. We determined a stellar 
radius of $R_\star$\,=\,\sradius.

We used the BAyesian STellar Algorithm \citep[\texttt{BASTA}, ][]{2015MNRAS.452.2127S} to determine a stellar mass of $M_*$\,=\smass\, and an 
age of $\tau_\star$\,=\,10\,$\pm$\,2~Gyr by fitting the stellar 
radius $R_*$, effective temperature \teff\ and iron abundance [Fe/H] 
to a large, finely-sampled grid of \texttt{GARSTEC} stellar models 
\citep{2008Ap&SS.316...99W}.

From the Ca~{\sc ii} H and K S-index values provided by the HARPS DRS, we calculated $\log R^{'}_{\rm HK}$\,=\,$-$5.07\,$\pm$\,0.03 \citep{Lovis2011}. Using 
the activity-rotation empirical relationships reported in \citet{Noyes1984} and \citet{Mamajek2008}, we derived a stellar rotation 
period of $P_{\rm rot}$\,=\,34\,$\pm$\,6 and 37\,$\pm$\,4 days respectively, which are in good mutual agreement. An upper limit 
to $P_{\rm rot}$ of 22$^{+13}_{-6}$ days can be inferred from the the stellar radius and \vsini\,, which is compatible with good alignment 
between the stellar rotation axis and the planetary orbital axis.
We note that the 27.9 day duration of the \tess\ observations is not long enough to 
attempt a reliable estimation of the photometric stellar rotational period.

\begin{table}
\caption{Fundamental  parameters and elemental abundances of HD\,219666.}
\label{table:starspecpar}
\centering
\begin{tabular}{l r c}
\hline\hline
\noalign{\smallskip}
Parameter & Value  \\
\hline
\noalign{\smallskip}
Star mass $M_{\star}$ [$M_\odot$]              & \smass[]        \\
\noalign{\smallskip}    
Star radius $R_{\star}$ [$R_\odot$]            & \sradius[]      \\
\noalign{\smallskip}    
Effective Temperature $\mathrm{T_{eff}}$ [K]   & 5527\,$\pm$\,65              \\
\noalign{\smallskip}    
Surface gravity \logg\ [cgs]                   & 4.40\,$\pm$\,0.11             \\
\noalign{\smallskip}    
Iron abundance [Fe/H] [dex]                    & 0.04\,$\pm$\,0.04           \\
\noalign{\smallskip}    
Project. rot. vel.               \vsini\ [\kms]   & 2.2\,$\pm$\,0.8             \\
\noalign{\smallskip}    
Micro-turb. vel.  $v_{\rm mic}$   [\kms]              & 0.9\,$\pm$\,0.1            \\
\noalign{\smallskip}    
Macro-turb. vel.  $v_{\rm mac}$  [\kms]               & 2.8\,$\pm$\,0.9            \\
\noalign{\smallskip}    
Ca {\sc ii} activity indicator $\log R^{'}_{\rm HK} $                         &  -5.07\,$\pm$\,0.03         \\
\noalign{\smallskip}    
Age $\tau_\star$ [Gyr]                                      & 10\,$\pm$\,2             \\
\noalign{\smallskip}    
Lithium abundance A(Li)           &      $<$0.40            \\
\hline
\noalign{\smallskip}
${\rm [C\,I/H]}$   &      0.074$\pm$0.065    \\
${\rm [O\,I/H]}$  &       0.043$\pm$0.148    \\
${\rm [Na\,I/H]}$   &     0.090$\pm$0.044    \\
${\rm [Mg\,I/H]}$   &     0.152$\pm$0.049    \\
${\rm [Al\,I/H]}$   &     0.196$\pm$0.041    \\
${\rm [Si\,I/H]}$   &     0.085$\pm$0.035    \\
${\rm [Ca\,I/H]}$   &     0.041$\pm$0.073    \\
${\rm [Sc\,II/H]}$  &     0.103$\pm$0.050    \\
${\rm [Ti\,I/H]}$   &     0.149$\pm$0.073    \\
${\rm [Ti\,II/H]}$  &     0.097$\pm$0.055    \\
${\rm [Cr\,I/H]}$   &     0.057$\pm$0.055    \\
${\rm [Ni\,I/H]}$   &     0.058$\pm$0.034    \\
${\rm [Cu\,I/H]}$   &     0.148$\pm$0.051    \\
${\rm [Zn\,I/H]}$   &     0.098$\pm$0.038    \\
${\rm [Sr\,I/H]}$   &    -0.034$\pm$0.105    \\
${\rm [Y\,II/H]}$   &     -0.057$\pm$0.057    \\
${\rm [Zr\,II/H]}$   &     0.027$\pm$0.073    \\
${\rm [Ba\,II/H]}$   &    -0.058$\pm$0.043    \\
${\rm [Ce\,II/H]}$   &     0.071$\pm$0.063    \\
${\rm [Nd\,II/H]}$   &     0.118$\pm$0.068    \\
${\rm [S\,I/H]}$   &      0.070$\pm$0.081    \\
\noalign{\smallskip}
\hline
\end{tabular}
\end{table}

\section{Joint analysis of the transit and Doppler data}\label{Sec:JointAnalysis}

We performed a joint fit to the \tess\ light curve (Sect.~\ref{Sec:Tess_Photometry}) and the 21 HARPS measurements (Sect.~\ref{Sec:spectroscopy}) using  
the code \texttt{pyaneti} \citep{Barragan2019}. The code uses a Bayesian approach for the model parameter estimations, and samples the posteriors 
via Markov chain Monte Carlo (MCMC) methods. 

We selected 10 hours of photometric data-points centred around each of the four transits observed by \tess\ and flattened the four segments 
using a second-order polynomial fitted to the out-of-transit data. We fitted the transit light curves using the limb-darkened quadratic model 
of \citet{Mandel2002}. We set Gaussian priors on the limb-darkening coefficients adopting the theoretical values predicted by \citet{Claret2017} along 
with a conservative error bar of 0.1 for both the linear and the quadratic limb-darkening term. The transit light curve poorly constrains the scaled 
semi-major axis ($a/R_\star$). We therefore set a Gaussian prior on $a/R_\star$ using the orbital period and the derived stellar 
parameters (Sect.~\ref{Sec:StellarParameters}) via Kepler's third law. 

The RV model consists of a Keplerian equation. Following \citep{Anderson2011}, we fitted for $\sqrt{e}\,\sin \omega_\star$ and $\sqrt{e}\,\cos \omega_\star$, where $e$ is 
the eccentricity and $\omega_\star$ is the argument of periastron. We also fitted for an RV jitter term to account for instrumental noise not included in the nominal 
uncertainties, and/or for RV variations induced by stellar activity. We imposed uniform priors for the remaining fitted parameters. Details of the fitted parameters and prior 
ranges are given in Table~\ref{tab:parameters}. 

\begin{figure}
\resizebox{\hsize}{!}{\includegraphics{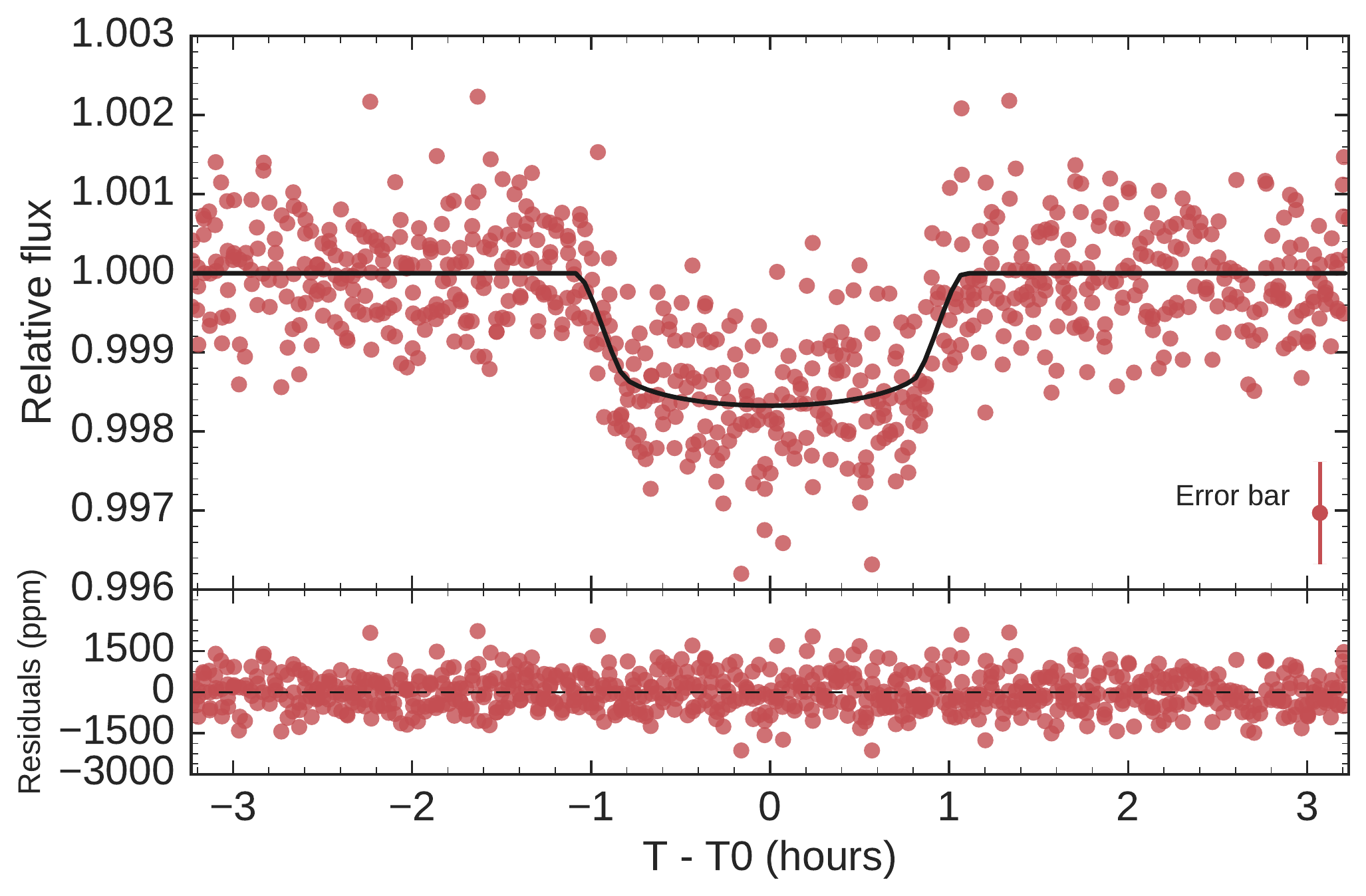}}
\caption{The phase-folded and normalized \tess\ photometric data with our
best fitting transit light curve.}
\label{fig:LCfit}
\end{figure}

\begin{figure}
\resizebox{\hsize}{!}{\includegraphics{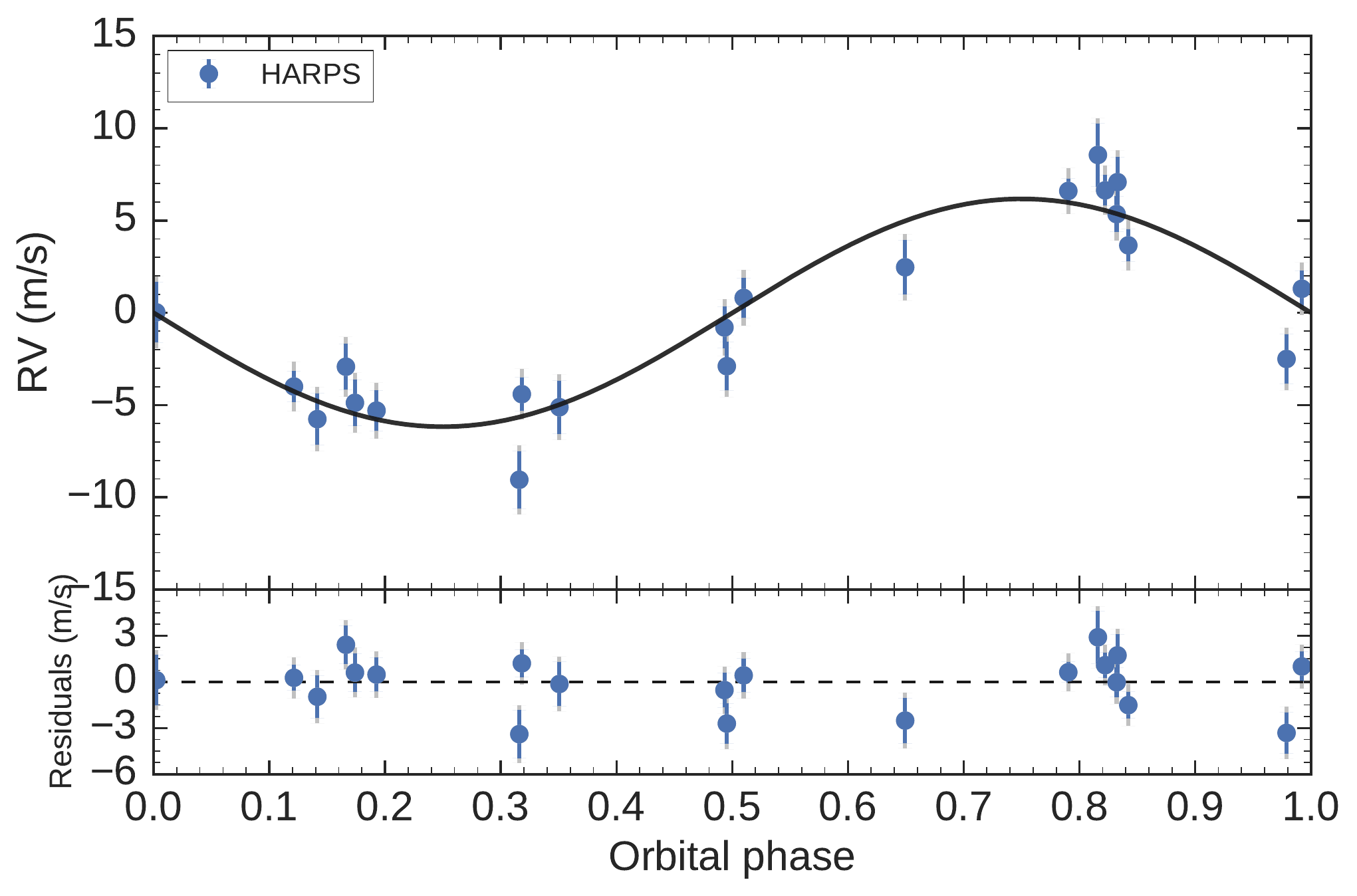}}
\caption{The phase-folded HARPS RV data points with our best fitting circular RV curve.
The blue error bars correspond to the nominal RV uncertainties calculated by the HARPS DRS while the grey
ones account for the RV jitter term. The bottom panel shows the RV residuals that have a rms of 1.7 \ms .  }
\label{fig:RVfit}
\end{figure}

We used 500 independent Markov chains initialized randomly inside the prior ranges. Once all chains converged, we used the last 5\,000 iterations and saved 
the chain states every ten iterations. This approach generates a posterior distribution of 250\,000 points for each fitted parameter. Table~\ref{tab:parameters} lists 
the inferred planetary parameters. They are defined as the median and 68\% region of the credible interval of the posterior distributions for each fitted parameter. The transit 
and RV curves are shown in Fig.~\ref{fig:LCfit} and \ref{fig:RVfit}, respectively.

An initial fit for an eccentric orbit yielded $e$\,=\,$0.07^{+0.06}_{-0.05}$, which is consistent with zero within less than 2$\sigma$. We determined the probability 
that the best-fitting eccentric solution could have arisen by chance if the orbit were actually circular using Monte Carlo simulations. Briefly, we created $10^5$ sets of 
synthetic RVs that sample the best-fitting circular solution at the epochs of our observations. We added Gaussian noise at the level of our measurements and 
fitted the simulated data allowing for an eccentric solution. We found that, given our measurements, there is a 35\,\% probability that an eccentric solution 
with $e\ge0.07$ could have arisen by chance if the orbit were actually circular. As this is above the 5\,\% significance level suggested by \citet{Lucy1971}, we decided 
to conservatively assume a circular model. We note that the eccentric solution provides a planetary mass that is consistent 
within less than 1-$\sigma$ of the result from the circular model.

\section{Discussion and conclusion}\label{Sec:Discussion}

\begin{figure}
\resizebox{1.02\hsize}{!}{\includegraphics{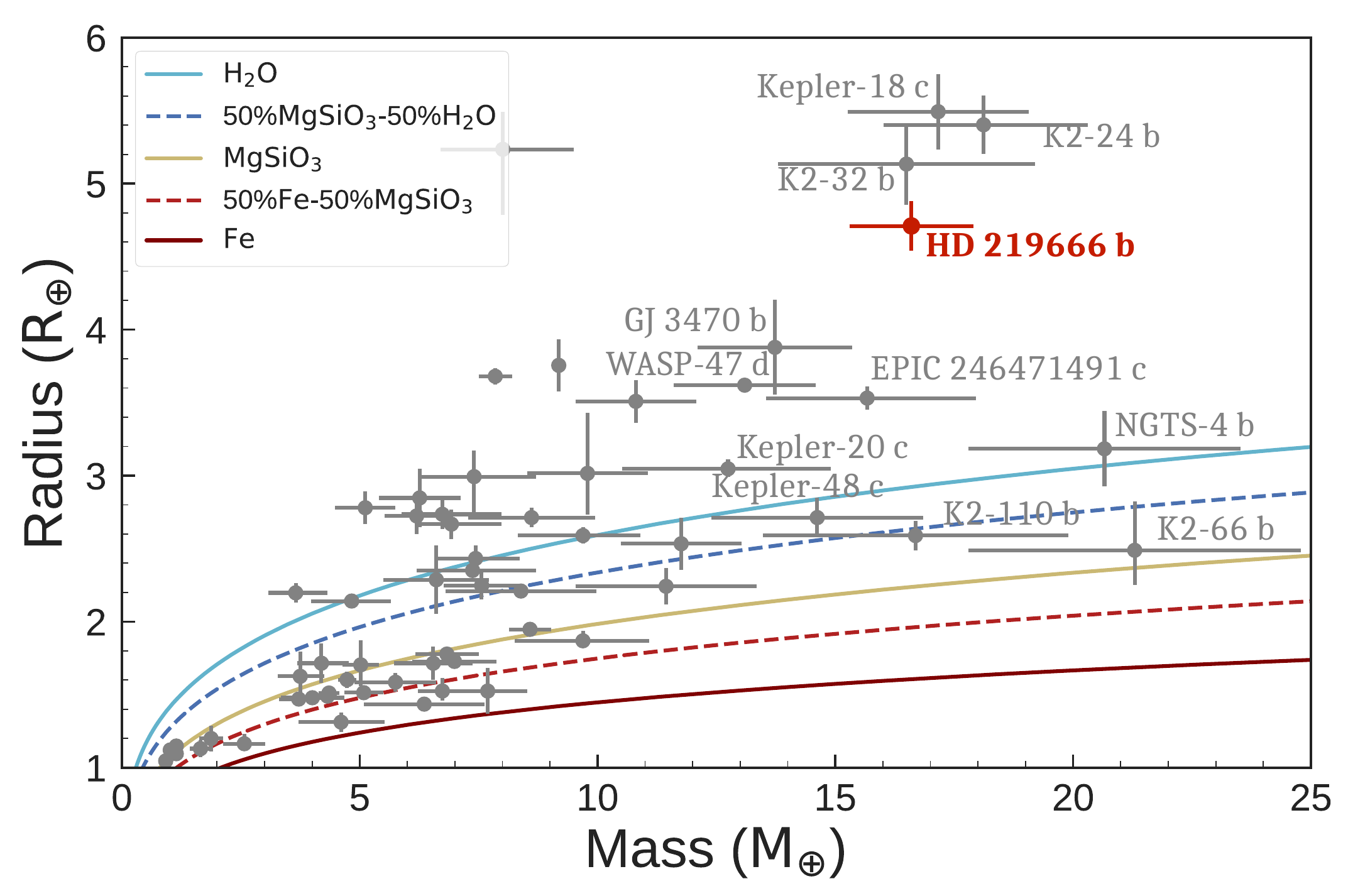}}
\caption{Mass$-$radius diagram for planets with masses $M_{\mathrm p}$\,$<$\,25~$M_\oplus$ and radii $R_{\mathrm p}$\,$<$\,6\,$R_\oplus$, as retrieved from the 
catalogue for transiting planets \texttt{TEPCat} \citep[available at \url{http://www.astro.keele.ac.uk/jkt/tepcat/};][]{Southworth2011}. Planets whose masses 
and radii are known with a precision better than 25\% are plotted with grey circles. Composition models from \citet{Zeng2016} are displayed with different lines 
and colours. The red circle marks the position of HD\,219666\,b. Planets closer in mass to HD\,219666\,b are labelled.}
\label{fig:plot_mr}
\end{figure}

HD\,219666\,b has almost the same mass as Neptune ($M_{\mathrm b}$\,=\,\mpb) but a  larger radius ($R_{\mathrm b}$\,=\,\rpb). With an orbital 
period of $P_\mathrm{orb}$\,$\simeq$\,6~days and an equilibrium temperature of $T_{\mathrm eq}\,\simeq$\,1073~K, it is a new member of a relatively 
rare class of exoplanets: the hot-Neptunes. Figure ~\ref{fig:plot_mr} shows that HD\,219666\,b lies in a region of the mass$-$radius diagram that is scarcely 
populated. The comparison with rocky planets composition models \citep{Zeng2016} suggests that HD\,219666\,b holds a conspicuous gas envelope. 

\begin{figure}
\resizebox{\hsize}{!}{\includegraphics{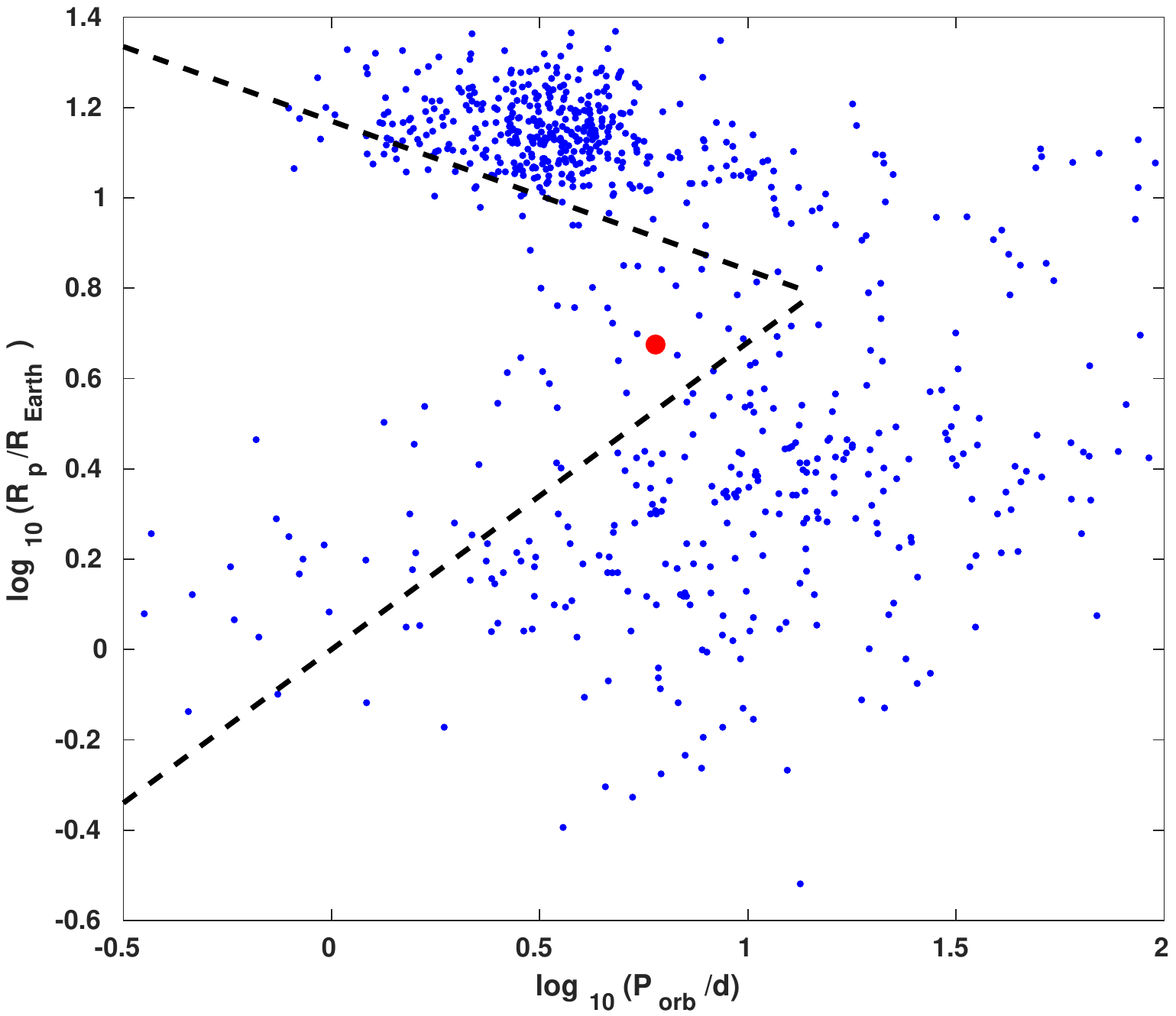}}
\caption{ Planet radius as a function of the orbital period. As for Fig. \ref{fig:plot_mr}, data are retrieved from \texttt{TEPCat}. 
The position of HD\,219666\,b is shown by a red circle.
The black dashed lines delimit the hot-Neptunes desert as derived in \citet{Mazeh2016}.}
\label{fig:plot_pr}
\end{figure}

The existence of a hot-Neptunes \textquotedblleft desert\textquotedblright \,has already been pointed out \citep[see, e.g.][]{Szabo2011,Mazeh2016,Owen2018}, and  
HD\,219666\,b falls close to the lower edge of the desert in the mass$-$period diagram \citep[see Fig. 1  in][]{Mazeh2016}, and well inside the desert in the radius$-$period 
diagram (see Fig. \ref{fig:plot_pr}). The relative paucity of hot-Neptunes (as compared to hot super-Earths and hot-Jupiters) could be interpreted 
as a consequence of two different formation mechanisms for short-period planets: in-situ formation for terrestrial planets \citep{Ogihara2018,Matsumoto2017}, and formation at 
larger separations followed by inward migration for giant planets \citep{Nelson2017}. Intermediate-mass planets like HD\,219666\,b would then be either the 
upper tail of terrestrial planets or the lower tail of giant-planet distributions. Alternatively, giant and small close-in planets could have a common origin 
but a dramatically different atmospheric escape history \citep{Lundkvist2016,Ionov2018,Owen2018}. 
 Other mechanisms have been proposed to explain the observed hot-Neptune desert. 
  \citet{Matsakos2016} advanced an explanation  based on high-eccentricity migration 
followed by tidal circularization. They interpreted the two distinct segments of the
desert boundary as a  consequence of the different slopes of the empirical mass–radius relation for small and large
planets. 
\citet{Batygin2016} advocated the in-situ formation of close-in super-Earths and hot-Jupiters alike. In the rare cases
when a core mass of $M_{\rm core}\gtrsim 15 M_\oplus$ was reached,  rapid gas accretion would lead to the formation of a gaseous giant
planet. In this way the relative occurrence of Earth- , Neptune- and Jupiter-like close-in planets can be explained.

To determine whether or not in-situ formation of a planet so close to its star is even possible, we calculate the isolation mass of a planet 
orbiting with a period of 6 days around a $0.9\, M_\odot$ star.  This is the mass of the planet that can form assuming that it grows by consuming 
all the planetesimals that are within its gravitational influence.  Assuming a typical T Tauri disc with a mass of $0.01 M_\odot$ within 100\,AU, a gas-to-dust 
ratio of 100 and a surface mass density profile of $\Sigma \propto R^{-3/2}$ (a steep profile enables as much material as possible to be made available in 
the inner disc for planet formation), the available rocky material is $\approx 5 \times 10^{-3}$\,$M_{\oplus}$. Even if the gas-to-dust ratio was a 
factor of ten lower \citep{Ansdell2016} the resulting mass is still nowhere near the mass of the planet reported in this study.  We point out that this calculation 
assumes no accretion through the disc when in reality rocky material could drift inwards and build up the core. From the perspective of 
pebble accretion, \citet{Lambrechts2014} showed that the pebble isolation mass -- the core mass at which the drift of pebbles ceases, stopping the accretion 
of rocky material onto the core -- at the radial location of the reported planet is approximately 1~$M_{\oplus}$, and simulations of planet growth by pebble 
accretion in evolving discs also show that 
 a high mass of rocky material cannot be produced in the inner discs \citep{Bitsch2015}. 
The $\sim$1~$M_{\oplus}$ upper limit to the rocky material that could have been accreted in situ  must be compared with an estimate of the $M_{\rm core}$ of HD\,219666\,b.
According to \citet{Lopez2014}, HD\,219666\,b, given its mass and radius,  should have a H/He envelope which contributes 10 to 20\% of its total mass, that is
the 80 to 90\% of the mass (13 to 15 $M_{\oplus}$) belongs to the rocky core.
Therefore, we conclude that it is more likely that HD\,219666\,b formed further out and migrated inwards.

We derived the atmospheric mass-loss rate of HD\,219666\,b using the interpolation routine presented by \citet{Kubyshkina2018}, which is based on a large 
grid of hydrodynamic upper atmosphere models. The main assumption is that the planet hosts a hydrogen-dominated atmosphere, which, given the measured bulk 
density, appears to be valid. For the computation, we employed the system parameters listed in Table \ref{table:starspecpar} and a high-energy 
stellar flux (hereafter referred to as XUV flux) at the planetary distance to the star of 573.8 erg\,cm$^{-2}$\,s$^{-1}$, obtained by scaling the solar 
XUV flux, derived from integrating the solar irradiance reference spectrum \citep{Woods2009} below 912~\AA, to the distance of the planet and 
the radius of the host star. This is a good assumption because the host star has a mass close to solar and appears to be rather inactive 
and old. We obtained a hydrogen mass-loss rate of about 1.2\,$\times$\,10$^{10}$~g\,s$^{-1}$, which is comparable to what is obtained employing the energy-limited 
formula \citep[5.2\,$\times$\,10$^9$~g\,s$^{-1}$;][]{Erkaev2007}. 
This indicates that, for this planet, atmospheric expansion and mass loss are driven mostly by atmospheric heating due to absorption of the stellar 
XUV flux, with an additional component due to the intrinsic thermal energy of the atmosphere and low planetary gravity \citep{Fossati2017}. The obtained 
mass-loss rate corresponds to 0.06 $M_{\oplus}$\,Gyr$^{-1}$, suggesting that mass loss does not play a major role in the current evolution of 
the planetary atmosphere. However, this does not account for the fact that the star was probably more active in the past, particularly during the first 
few hundred million years, up to about 1~Gyr \citep{Jackson2012,Tu2015}, when the XUV fluxes could have been up to about 500 times larger than the current 
estimate. This would lead to mass-loss rates about 500 times higher. It is therefore likely that atmospheric escape played a significant role in 
shaping the early planetary atmospheric evolution.

HD 219666 b is an interesting target for further atmospheric
characterisation, given its equilibrium temperature of $\sim$1070~K,
since the range of expected temperatures at the terminator (depending
on the planet's albedo and energy transport) straddles widely
different atmospheric chemical compositions under thermochemical
equilibrium.
Using the properties of the system, we modelled the transmission 
spectrum of the planet using the Python Radiative Transfer in a Bayesian framework\footnote{\href{http://pcubillos.github.io/pyratbay}{http://pcubillos.github.io/pyratbay}} (Cubillos et al., in prep.), which 
is based on the Bayesian Atmospheric Radiative Transfer package \citep{Blecic2016phdThesis, Cubillos2016phdThesis}, and simulated {\it James Webb Space 
Telescope (JWST)} observations with Pandexo \citep{BatalhaEtal2017paspPandexo}.
These models consider opacities from the main spectroscopically active species expected for exoplanets at these 
wavelengths: {\water} and {\carbdiox} from \citet{RothmanEtal2010jqsrtHITEMP}; {\methane}, {\ammonia}, and HCN 
from \citet{YurchenkoTennyson2014mnrasExomolCH4}; CO from \citep{LiEtal2015apjsCOlineList}; Na and K from \citet{BurrowsEtal2000apjBDspectra}; Rayleigh opacities 
from H, He, and {\molhyd} \citep{Kurucz1970saorsAtlas, LecavelierDesEtangsEtal2008aaRayleighHD189}; and collision-induced absorption 
from {\molhyd}--{\molhyd} \citep{BorysowEtal2001jqsrtH2H2highT,Borysow2002jqsrtH2H2lowT} and {\molhyd}--He \citep{BorysowEtal1988apjH2HeRT,borysowfrommhold1989b, borysowfrommhold1989a}. We 
compressed the HITEMP and ExoMol databases with the open-source repack package \citep{Cubillos2017apjCompress} to extract only the strong, dominating line transitions.

\begin{figure}
\resizebox{\hsize}{!}{\includegraphics{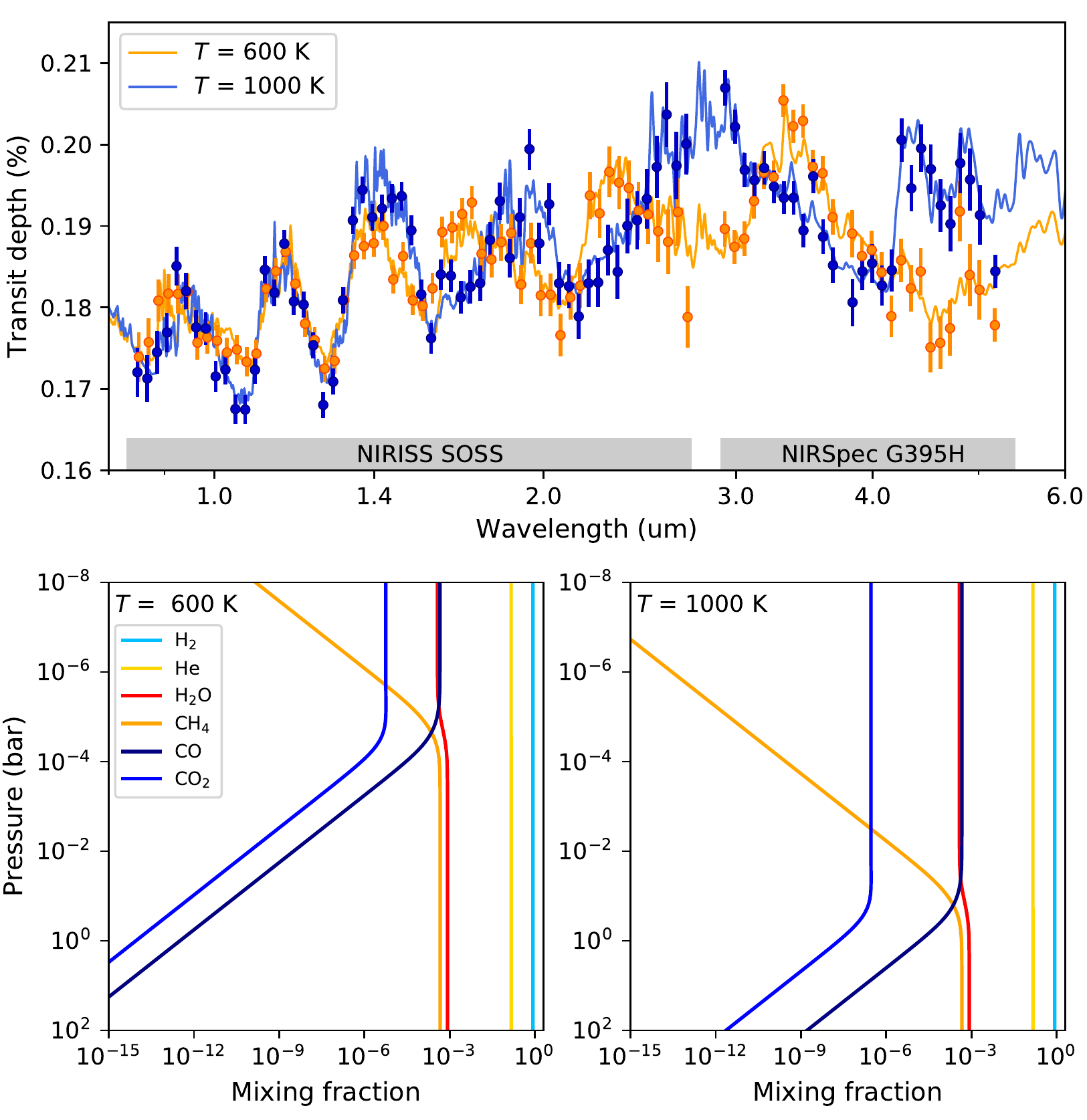}}
\caption{Model transmission spectra of HD 219666 b (top panel).  The dots and error bars denote simulated single-transit {\it JWST} transmission 
observations with NIRISS SOSS and NIRSpec G395H (wavelength coverage at bottom) for two underlying models (solid curves) at temperatures of 
600~K and 1000~K (see legend).  {\methane} shows strong absorption bands at 1.7, 2.3, and 3.3 $\mu$m in the 600~K model; whereas 
CO and {\carbdiox} show their strongest absorption features at wavelengths beyond 4 $\mu$m in the 1000~K model.  The bottom panels show the 
composition of the main species that shape the transmission spectrum. 
 Depending on the atmospheric temperature, carbon favours either higher
{\methane} (temperatures lower than $\sim$900~K) or CO/{\carbdiox} abundances
(otherwise).
}
\label{fig:JWSTsim}
\end{figure}

Figure~\ref{fig:JWSTsim} shows estimated transmission spectra of HD\,219666\,b assuming a cloud-free atmosphere, in thermochemical equilibrium \citep{BlecicEtal2016apsjTEA} for 
solar elemental composition, 
 at two illustrative atmospheric temperatures that lead
to different transmission spectra.
By combining NIRISS SOSS and NIRSpec G395H observations, one could potentially constrain 
the atmospheric chemistry and temperature of the planet with a single-transit observation with each instrument.  The transmission spectrum at wavelengths shorter 
than 2~$\mu$m constrain the {\water} abundance for both models, setting the baseline to constrain the abundances of other species. At longer 
wavelengths, either {\methane} ($T\,=\,600$~K model) or CO/{\carbdiox} ($T\,=\,1000$~K model) dominate the carbon chemistry at the probed 
altitudes (Fig.~\ref{fig:JWSTsim}, bottom panels), producing widely different features in the transmission spectrum (Fig.~\ref{fig:JWSTsim}, top panel).

An important clue to the formation mechanism of HD\,219666\,b could come from the knowledge of its orbital obliquity with respect 
to the stellar equatorial plane, which can be estimated through the observation of the Rossiter-McLaughlin (RM) effect. 
We calculated that the RV amplitude of the RM effect is of $\sim$3~\ms, meaning that it would probably be detectable with HARPS, and certainly 
with ESPRESSO \citep{Pepe2010}. Remarkably, there are only two hot-Neptunes with a reported measure of the orbital 
obliquity, \object{GJ\,436\,b} \citep{Bourrier2018} and \object{HAT-P-11\,b}
\citep{Winn2010}, and both have a misaligned orbit . 

Given the precise RV measurements from HARPS and the mid-transit time from the TESS mission, we can also constrain the presence 
of co-orbital planets (or trojans) to HD\,219666\,b, by putting upper limits to their mass $M_{\rm t}$ (assuming there are no other planets in the 
system or they are far enough to not perturb the RVs in the time span of our observations). We followed the technique described in \citet{Leleu2017}, and 
subsequently applied in \citet{LilloBox2018a,LilloBox2018b}, to model the RV data by including the so-called $\alpha$ parameter, which accounts for 
the possible mass imbalance between the L$_4$ and L$_5$ regions in the co-orbital region of the planet. The parameter $\alpha$ is defined 
as $M_{\rm t}/M_{\rm b}\sin{\theta}+ \mathcal{O}(e^2)$, where $\theta$  is the resonant angle representing the difference between the mean 
longitudes of the trojan and the planet. We set Gaussian priors on the time of transit and period of the planet, and left the 
rest of the parameters (i.e. $e\cos{\omega}$, $e\sin{\omega}$, $\alpha$, $\gamma$, and $K_{\rm b}$) with uniform broad priors. We also 
included a slope term and a jitter term to account for white noise. The result of this analysis provides parameters compatible with the prior 
joint analysis and allows us to set constraints on co-orbital planets in the system. In particular, we find $\alpha=-0.14\pm0.22$, which assuming 
the estimated planet mass provides an upper limit (95\% confidence level) of $M_{\rm t}$\,=\,4.6~$M_{\oplus}$ at L5 and no constraint (i.e. up to the mass of the planet) at L4.

In conclusion, we report the discovery of a hot-Neptune transiting the bright (V=9.9) G7\,V star HD\,219666. The collaboration between 
the \texttt{KESPRINT} and \texttt{NCORES} consortia has made possible a rapid spectroscopic follow-up with HARPS, leading to the confirmation 
and characterisation of the planet candidate detected by \tess. HD\,219666\,b adds to a list of only five Neptune-like 
planets (0.5\,<\,$M_{\rm p}$\,<\,2~$M_{\rm Nep}$ with 1~M$_{\rm Nep}$\,=\,17.2\,$M_\oplus$) transiting a V\,$<$\,10 star. We carried out detailed 
analyses to derive the fundamental parameters and the elemental abundances of the host star. We discuss the possibility of further characterisation 
of the planet, in particular by examining the potential of {\it JWST}  in-transit observations to detect the presence of molecular features in transmission spectra.

\begin{table*}
\begin{center}
\caption{\label{Table:HARPS} HARPS RV measurements of \sname.}
\begin{tabular}{lcccccccc}
\hline\hline
\noalign{\smallskip}
$\rm BJD_{TDB}^a$ & RV & $\sigma_\mathrm{RV}$ &  BIS  & FWHM    & S-index &
$\sigma_\mathrm{S-index}$ & T$_\mathrm{exp}$ & S/N$^b$\\
-2450000          & [\kms]  & [\kms]     &  [\kms] & [\kms]  &  &  & [s] &   \\
\hline
\noalign{\smallskip}
\noalign{\smallskip}
 8394.521096 & -20.0909 & 0.0008 & -0.0274 & 6.9061 & 0.154 & 0.001 & 1200 &  87.9 \\
 8394.641680 & -20.0939 & 0.0009 & -0.0281 & 6.9033 & 0.152 & 0.002 & 1200 &  85.7 \\
 8396.644285 & -20.1024 & 0.0012 & -0.0242 & 6.9048 & 0.144 & 0.003 & 1200 &  62.5 \\
 8396.756848 & -20.1029 & 0.0011 & -0.0267 & 6.9081 & 0.147 & 0.003 & 1200 &  72.4 \\
 8397.501496 & -20.1066 & 0.0016 & -0.0274 & 6.9102 & 0.143 & 0.004 & 1500 &  50.9 \\
 8397.710686 & -20.1027 & 0.0014 & -0.0253 & 6.9070 & 0.144 & 0.003 & 1200 &  54.2 \\
 8398.571357 & -20.0984 & 0.0011 & -0.0278 & 6.9130 & 0.148 & 0.002 & 1200 &  67.3 \\
 8398.671630 & -20.0968 & 0.0011 & -0.0264 & 6.9103 & 0.144 & 0.002 & 1200 &  70.3 \\
 8399.513841 & -20.0951 & 0.0015 & -0.0316 & 6.9114 & 0.148 & 0.004 & 1200 &  53.4 \\
 8401.643664 & -20.0975 & 0.0016 & -0.0280 & 6.9094 & 0.161 & 0.006 & 1200 &  51.8 \\
 8404.619501 & -20.1005 & 0.0013 & -0.0295 & 6.9103 & 0.145 & 0.003 & 1200 &  59.9 \\
 8406.554873 & -20.0890 & 0.0017 & -0.0282 & 6.9095 & 0.146 & 0.004 & 1200 &  47.3 \\
 8406.657043 & -20.0905 & 0.0014 & -0.0225 & 6.9092 & 0.140 & 0.003 & 1200 &  58.1 \\
 8407.538610 & -20.1001 & 0.0013 & -0.0242 & 6.9058 & 0.144 & 0.003 & 1200 &  58.0 \\
 8407.618837 & -20.0963 & 0.0010 & -0.0274 & 6.9078 & 0.150 & 0.002 & 1200 &  78.3 \\
 8408.519940 & -20.1033 & 0.0014 & -0.0304 & 6.9129 & 0.153 & 0.003 & 1200 &  55.6 \\
 8408.668982 & -20.1005 & 0.0012 & -0.0285 & 6.9096 & 0.145 & 0.003 & 1200 &  64.8 \\
 8424.508079 & -20.0910 & 0.0007 & -0.0263 & 6.9102 & 0.153 & 0.001 & 1200 & 108.7 \\
 8424.760122 & -20.0922 & 0.0010 & -0.0262 & 6.9148 & 0.144 & 0.003 & 1200 &  84.8 \\
 8426.505548 & -20.1016 & 0.0008 & -0.0236 & 6.9117 & 0.153 & 0.001 & 1200 &  86.6 \\
 8427.693940 & -20.1020 & 0.0009 & -0.0267 & 6.9079 & 0.152 & 0.002 & 1200 &  89.2 \\
\noalign{\smallskip}
\hline
\end{tabular}
\\
\end{center}
Notes:\\
$^a$ Barycentric Julian dates are given in barycentric dynamical time. \\
$^b$ S/N per pixel at 5500\,\AA. \\
\end{table*}

\begin{table*}[!t]
\begin{center}
  \caption{\sname\ system parameters. \label{tab:parameters}}  
  \begin{tabular}{lcc}
  \hline\hline
  \noalign{\smallskip}
  Parameter & Prior$^{(\mathrm{a})}$ & Derived value \\
  \noalign{\smallskip}
  \hline
    \noalign{\smallskip}
    \multicolumn{3}{l}{\emph{\bf{Model parameters of \pname}}} \\
    \noalign{\smallskip}
    Orbital period $P_{\mathrm{orb}, \mathrm{b}}$ (days) &  $\mathcal{U}[ 6.00 , 6.08]$ &
\Pb[] \\
    \noalign{\smallskip}
    Transit epoch $T_\mathrm{0, \mathrm{b}}$ (BJD$_\mathrm{TDB}-$2\,450\,000) &
$\mathcal{U}[8329.10 , 8329.30]$ & \Tzerob[]  \\ 
    \noalign{\smallskip}
    Scaled semi-major axis $a_\mathrm{b}/R_{\star}$ &  $\mathcal{N}[14.39,0.30]$ &
\arb[] \\
    \noalign{\smallskip}
    Planet-to-star radius ratio $R_\mathrm{b}/R_{\star}$ & $\mathcal{U}[0,0.1]$ &
\rrb[]  \\
    \noalign{\smallskip}
    Impact parameter $b_\mathrm{b}$  & $\mathcal{U}[0,1]$  & \bb[] \\
    \noalign{\smallskip}
    $\sqrt{e} \sin \omega_\mathrm{\star}$ &  $\mathcal{F}[0]$  & 0 \\
    $\sqrt{e} \cos \omega_\mathrm{\star}$ &  $\mathcal{F}[0]$  & 0 \\
    Radial velocity semi-amplitude variation $K_\star$ (\ms) &
$\mathcal{U}[0,10]$ & \kb[] \\
    \noalign{\smallskip}
    \hline    
    \noalign{\smallskip}
    \multicolumn{3}{l}{\emph{\bf{Additional model parameters}}} \\
    \noalign{\smallskip}
    Parameterized limb-darkening coefficient $q_1$  & $\mathcal{N}[0.34,0.1]$ &
\qone \\
    \noalign{\smallskip}    
    Parameterized limb-darkening coefficient $q_2$  & $\mathcal{N}[0.23,0.1]$ &
\qtwo \\
    Systemic velocity $\gamma_{\mathrm{HARPS}}$  (km s$^{-1}$) & $\mathcal{U}[-20.30
, -19.9]$ & \HARPS[]  \\
   \noalign{\smallskip}
    RV jitter term $\sigma_{\mathrm{HARPS}}$  (\ms) & $\mathcal{U}[0,100]$ &
\jHARPS[]  \\
   \noalign{\smallskip}
    \hline
    \noalign{\smallskip}
    \multicolumn{3}{l}{\emph{\bf{Derived parameters of \pname}}} \\
    \noalign{\smallskip}

    Planet mass $M_\mathrm{b}$ ($M_{\oplus}$) & $\cdots$ & \mpb[]  \\
    \noalign{\smallskip}
    Planet radius $R_\mathrm{b}$ ($R_{\oplus}$) & $\cdots$ & \rpb[] \\
    \noalign{\smallskip}
    Planet mean density $\rho_\mathrm{b}$ ($\mathrm{g\,cm^{-3}}$) & $\cdots$ &
\denpb[] \\
    \noalign{\smallskip}    
    Semi-major axis of the planetary orbit $a_\mathrm{b}$ (AU) & $\cdots$ & \ab[]  \\
    \noalign{\smallskip}
    Orbit eccentricity $e_\mathrm{b}$ & $\cdots$ & 0 (fixed) \\
    \noalign{\smallskip}
    Orbit inclination $i_\mathrm{b}$ (deg) & $\cdots$ & \ib[] \\
    \noalign{\smallskip}
    Equilibrium temperature$^{(\mathrm{d})}$  $T_\mathrm{eq,\,b}$ (K)  & $\cdots$ & 
\Teqb[] \\
    \noalign{\smallskip}
    Transit duration $\tau_\mathrm{14,\,b}$ (hours) & $\cdots$ & \ttotb[] \\
    \noalign{\smallskip}
  \hline
  \end{tabular}
  \begin{tablenotes}\footnotesize
  \item \emph{Note} -- 
                       $^{(\mathrm{a})}$ $\mathcal{U}[a,b]$ refers to uniform priors between $a$ and $b$, and $\mathcal{F}[a]$ to a fixed $a$ value.
                       $^{(\mathrm{b})}$ From spectroscopy and isochrones.
                       $^{(\mathrm{c})}$ From spectroscopy.
                       $^{(\mathrm{d})}$ Assuming zero albedo and uniform redistribution of heat.
\end{tablenotes}
\end{center}
\end{table*}

\begin{acknowledgements}
We thank the anonymous referee for her/his useful comments.
This  paper  includes  data  collected  by  the  TESS  mission,
which are publicly available from the Mikulski Archive for Space Telescopes
(MAST).  Funding  for  the  TESS  mission  is  provided  by  NASA’s  Science
Mission directorate.
ME acknowledges the support of the DFG priority program SPP 1992
"Exploring the Diversity of Extrasolar Planets" (HA 3279/12-1).
DB acknowledges support by the Spanish State Research Agency (AEI) Project No. ESP2017-87676-C5-1-R and No. MDM-2017-0737 Unidad 
de Excelencia "María de Maeztu" - Centro de Astrobiología (INTA-CSIC). DJA gratefully acknowledges support from the STFC via 
an Ernest Rutherford Fellowship (ST/R00384X/1). This work was supported by FCT - Fundação para a Ciência e a Tecnologia through national 
funds and by FEDER through COMPETE2020 - Programa Operacional Competitividade e Internacionalização by 
these grants: UID/FIS/04434/2013 \& POCI-01-0145-FEDER-007672; PTDC/FIS-AST/28953/2017 \& POCI-01-0145-FEDER-028953 and PTDC/FIS-AST/32113/2017 \& POCI-01-0145-FEDER-032113. X.D is grateful 
to the Branco-Weiss fellowship $-$ Society in Science for its financial support. PJW is supported by the STFC Consolidated Grant ST/P000495/1. S.H. acknowledges 
support by the fellowships PD/BD/128119/2016 funded by FCT (Portugal). 
This work has made use of the VALD database, operated at Uppsala University, the Institute of Astronomy RAS in Moscow, and the University of Vienna.
This publication makes use of The Data \& Analysis Center for Exoplanets (DACE), which is a facility based at the University of Geneva (CH) dedicated to 
extrasolar planets data visualisation, exchange and analysis. DACE is a platform of the Swiss National Centre of Competence in Research (NCCR) PlanetS, federating 
the Swiss expertise in Exoplanet research. The DACE platform is available at \url{https://dace.unige.ch}. We  thank  the  Swiss  National  Science  Foundation  (SNSF) and 
the Geneva University for their continuous support to our planet search programs. This work has been in particular carried out in the frame of the National Centre 
for Competence in Research "PlanetS" supported by the Swiss National Science Foundation (SNSF).
S.C.C.B., N.C.S., S.G.S, V.A. and E.D.M. acknowledge support from FCT through Investigador FCT 
contracts nr. IF/01312/2014/CP1215/CT0004, IF/00169/2012/CP0150/CT0002, IF/00028/2014/CP1215/CT0002, IF/00650/2015/CP1273/CT0001, and IF/00849/2015/CP1273/CT0003.
This work is partly supported by JSPS KAKENHI Grant Numbers JP18H01265 and 18H05439,
and JST PRESTO Grant Number JPMJPR1775.
PGB acknowledges the support of the MINECO under the fellowship program 'Juan de
la Cierva incorporacion' (IJCI-2015-26034).
SM acknowledges support from the Ramon y Cajal fellowship number
RYC-2015-17697.
\end{acknowledgements}

\bibliographystyle{aa} 
\bibliography{ref_toi118.bib} 


\listofobjects

\end{document}